# Two – parameter scaling of conductance in quantum Hall effect

*Yu.G. Arapov, S.V. Gudina, V.N. Neverov, N.S. Sandakov, N.G. Shelushinina*

M.N. Mikheev lnstitute of Metal Physics of Ural Branch of Russian Academy of Sciences (IMP UB RAS), 18 S. Kovalevskaya Street, Yekaterinburg, Russian Federation, 620108

**Abstract**

After a brief survey of theoretical concepts for the two-parameter scaling theory in the integer quantum Hall effect (IQHE) regime, a comprehensive set of early, recent and new experimental results on constructing scaling diagrams for conductance in 2D semiconductor structures, as well as in graphene is displayed. A comparative analysis of scaling diagrams obtained from experimental data with calculated ones is carried out.

## *Content*

1. Introduction

2. Two – parameter scaling theory of IQHE

3. Scaling diagrams of conductance (theory)

4. Scaling diagrams (experiment)

5. Discussion and conclusions

References

## 1. Introduction

The integer quantum Hall effect (IQHE) [1] is a universal phenomenon that occurs when two-dimensional (2D) semiconductor systems are exposed to a perpendicular magnetic field at low temperatures. The magnetic field splits the constant density of states into discrete Landau levels, which are separated by energy gaps $\Delta E$. When the Fermi energy $E_F$ is in the gapped regions, the Hall conductance is quantized to integer multiples of $e^2/h$, where $h$ is Planck's constant and $e$ is the electron charge.

The IQHE manifests itself as a stair-like sequence of plateaus in the Hall magnetoresistivity (MR), $\rho_{xy}(B)$, located at its monotonically increasing values $\rho_{xy}(B) = h/ie^2$ for integer filling factors $i$, with zero longitudinal MR, $\rho_{xx}(B)$, in the plateau regions [2, 3].

The origin of the QHE can be explained on the basis of localized and extended states that arise in the spectrum of impurity-broadened Landau levels [4, 5]. At the Landau level centers there are extended states that lead to metallic behavior, whereas in the intervals of localized states the bulk behavior is insulating.

The existence of localized states in semiconductors has been known since the seminal work of Anderson [6, 7]. He used a simple in appearance tight binding model and showed that when the disorder of the binding energy exceeds a given limit, the wave functions are localized in space. I.e. the eigenstates fall off exponentially while still forming a continuous energy spectrum. The physics behind this mechanism is destructive interference.

Although bulk states in between Landau levels are localized, dissipationless 1D edge channels are formed that dominate the transverse transport properties in this regime [8,9]. The consequence is a quantized Hall resistance accompanied by a vanishingly small longitudinal resistance (see reviews [10,11]).

Thus, the localization-delocalization phenomenon is closely related to the very existence of the QHE. One of the possible approaches to understanding the QHE is to view the QHE regime as a series of localization - delocalization - localization (insulator-metal-insulator) quantum phase transitions in a quantizing magnetic field. In this case, it is possible to use previously developed methods for examining metal-dielectric transitions (see, for example, the monograph by Patashinsky and Pokrovsky [12] and the comprehensive review by Sadovsky [13]).

In particular, the application of the renormalization group method (scaling hypothesis) proposed by Abrahams et al. in their famous work [14] for disorder-controlled metal-insulator transition has proven quite effective. The variant of two-parameter scaling theory for IQHE, proposed by Pruisken [15] as a generalization of the scaling hypothesis to the case of a 2D system in a strong magnetic field, made it possible to describe in a single system of equations the existence of both localized (in the plateau regions of the QHE) and extended (for the plateau-plateau transitions) states.

The most widely used version of the scaling hypothesis focuses on the consideration of the quantum Hall plateau transition as a special case of an insulator-metal-insulator transition with the divergence of the localization length at the critical point in the Landau level center (see [16] and references therein).

In experiments the data on the magnetic-field and temperature dependences of resistivity components in the regions of the plateau-plateau transitions (PPTs) are analyzed within framework of a scaling hypothesis. After a brief presentation of the theoretical concepts of the two-parameter scaling hypothesis for the IQHE in Sections 2 and 3 of this work, in the main Section 4 we give a survey on literature examples of constructing scaling diagrams in coordinates $(\sigma_{xy}, \sigma_{xx})$ based on experimental data for various 2D systems. The extent to which the experimental graphs correspond (or do not correspond) to the ideal calculated ones is than discussed.

## 2. Two – parameter scaling theory of IQHE

The phenomenon of the integer quantum Hall effect, detected by Von Klitzing *et al.* [1], is closely associated with the problem of electron localization in a two- dimensional (2D) system in a quantizing magnetic field *B*. Laughlin [4] and Halperin [8] showed that, for the IQHE to exist, narrow bands of delocalized states must be present close to the middle of each of the Landau subbands (provided that all the other states are localized). On the other hand, in earlier work, for *B* = 0, Abrahams *et al.* [14] used the theory of one- parameter scaling to conclude that quantum diffusion is absent in 2D disordered systems; i.e., there are no delocalized states in 2D systems in the presence of even a small degree of disorder. The conclusions of Laughlin [4] and Halperin [8] thus contradicted the consequences of the theory of single-parameter scaling of Ref. 14.

To explain the IQHE, Pruisken [15, 17, 18] and also Khmel'nitskii [19] proposed the hypothesis of two-parameter scaling, which results in the existence of both localized and delocalized states (close to the middle of the Landau subbands) in the spectrum of a disordered *2D* system in a quantizing magnetic field. The consequences of the theory of two- parameter scaling in the IQHE regime were experimentally verified in early works of Wei *et al.* [20] for InGaAs/lnP heterostructures, of Kawaji *et al.* [21] for *n-* channels in a silicon MOS transistor, and of Dolgopolov *et al.* [22] for silicon MOS structures and for AlGaAs/GaAs heterostructures.

According to the hypothesis of single-parameter scaling [14, 23] the variation of the conductance (the inverse total resistance) *G* as the macroscopic size *L* of the system varies is determined by:

$$\frac{d \ln g}{d \ln L} = \beta(g) \qquad (1)$$

where $g=hG/e^2$, and $\beta$ is a function of the single variable $g$ (scaling function). If the conductance is represented as $g = \sigma L^{d-2}$, where $\sigma$ is the conductivity in units of $e^2/h$, and $d$ is the dimension, metallic behavior of the system corresponds to the condition $\sigma$=const as $L\to\infty$. For 2D systems, the concepts of conductance and conductivity coincide, and the condition of delocalization of the electronic states corresponds to the condition $\beta(g) = 0$.

As shown in Refs. 14 and 23, when $B=0$, for a $2D$ gas, the function $\beta(g)$ is always negative and only when $\sigma \to \infty$, which corresponds to the absence of disorder, does it asymptotically tend to zero. When $\sigma \gg 1$ it is:

$$\beta(g) = -1/2\pi\sigma. \tag{2}$$

Thus, for an electron in a disordered *2D* system, no true delocalized states (states with an infinite localization radius) exist. In the framework of the hypothesis of single parameter scaling, for dissipative conductivity $\sigma_{xx}$, the conclusion that all the states in an infinite 2D system are localized is maintained even in a magnetic field; i.e., $\beta(\sigma_{xx}) < 0$ for all finite values of $\sigma_{xx}$ [15, 18]. In the limit $\sigma_{xx} \gg 1$, the scaling function has the form:

$$\beta(\sigma_{xx}) = -1/2\pi^2\sigma_{xx}^2. \tag{3}$$

Pruisken [17] was the first to express the idea that, in a quantizing magnetic field, it is necessary to consider renormalization (with variation of *L)* of both components of the conductivity tensor: the dissipative component $\sigma_{xx}$ and the Hall component $\sigma_{xy}$. As a result, for noninteracting electrons in a chaotic (random) impurity (disorder) potential, we have the following system of equations of two-parameter scaling (see [15, 18]):

$$\frac{d \ln \sigma_{xx}}{d \ln L} = \beta_{xx}(\sigma_{xx}, \sigma_{xy}),$$
$$\frac{d \ln \sigma_{xy}}{d \ln L} = \beta_{xy}(\sigma_{xx}, \sigma_{xy}) \tag{4}$$

Each of the scaling functions $\beta_{xx}$ and $\beta_{xy}$ is a function of the two parameters $\sigma_{xx}$ and $\sigma_{xy}$; i.e., the variations of $\sigma_{xx}$ and $\sigma_{xy}$ are mutually dependent when *L* is varied. In the weak-localization limit $\sigma_{xx} \gg 1$, Pruisken obtained the specific form of the scaling functions:

$$\beta_{xx} = -\frac{1}{2\pi^2\sigma_{xx}^2} - D \exp(-2\pi\sigma_{xx}) \cos 2\pi\sigma_{xy},$$
$$\beta_{xy} = -D \exp(-2\pi\sigma_{xx}) \sin 2\pi\sigma_{xy}, \tag{5}$$

where $\sigma_{xx}$ and $\sigma_{xy}$ are given in units of $e^2/h$, and $D$ is a positive constant that contains information on the microscopic behavior of the system (for example, on the character of the chaotic impurity (disorder) potential).

It is convenient to study the consequences of the scaling equations by considering the motion of the points on the $(\sigma_{xx}, \sigma_{xy})$ plane as $L$ increases (scaling diagrams). It directly follows from Eqs. (4) and (5) that two types of fixed points exist on a phase diagram in $(\sigma_{xx}, \sigma_{xy})$ coordinates.

- When $\sigma_{xy} = i$, where $i$ is an integer, we have

$$\beta_{xy} = 0, \quad \beta_{xx} = -\frac{1}{2\pi^2 \sigma_{xx}^2} - D\exp(-2\pi\sigma_{xx}) < 0 \qquad (6)$$

Consequently, $\sigma_{xy}$ does not change as $L$ varies, while $\sigma_{xx} \to 0$ as $L \to \infty$, which indicates that the sample behaves like an insulator. Thus, for an infinite sample, we have a fixed point $(i, 0)$ that describes a plateau of the QHE.

- When $\sigma_{xy} = i + 1/2$, we have

$$\beta_{xy} = 0, \quad \beta_{xx} = -\frac{1}{2\pi^2 \sigma_{xx}^2} + D\exp(-2\pi\sigma_{xx}) \qquad (7)$$

It thus follows that $\sigma_{xy}$, as in the preceding case, is not renormalized as $L$ varies. It is further assumed that there is a finite value of $\sigma_{xx} = \sigma^*$, which is determined by the condition

$$(\sigma^*)^2 \exp(-2\pi\sigma^*) = \frac{1}{2\pi^2 D} \qquad (8)$$

for which $\beta_{xx} = 0$. In this case, $\beta_{xx} < 0$ for $\sigma_{xx} > \sigma^*$ and $\beta_{xx} > 0$ for $\sigma_{xx} < \sigma^*$. Then, as $L \to \infty$, we have $\sigma_{xx} \to \sigma^*$ (metallic behavior), and the fixed point $(i + 1/2, \sigma^*)$ corresponds to a delocalized state in the center of the Landau subband.

By extrapolating Eqs. (4) into the region of the strong localization regime $\sigma_{xx} \ll 1$, Pruisken [15, 18] obtained the qualitative form of the scaling diagrams on the $(\sigma_{xy}, \sigma_{xx})$ plane, i.e., the form of the trajectories of the matched transformation of the conductivities $\sigma_{xx}$ and $\sigma_{xy}$ as $L$ varies from a value of the order of the magnetic length to $L \to \infty$ (see Fig. 1) These diagrams have been repeatedly reproduced in the literature (see, for example, Refs. 18-20, 22).

## 3. Scaling diagrams of conductance (theory)

In Fig. 1, we show a section of the scaling diagram for one Landau level (the zeroth level) from Ref. 18. The main features of the scaling diagrams are as follows: the presence of two types of fixed points, as well as the existence of a universal trajectory (a separatrix), along which points on the $(\sigma_{xy}, \sigma_{xx})$ plane "flow" from the neighborhood of the point $(i + 1/2, \sigma^*)$ to the points $(i, 0)$ and $(i + 1, 0)$ as $L$ increases.

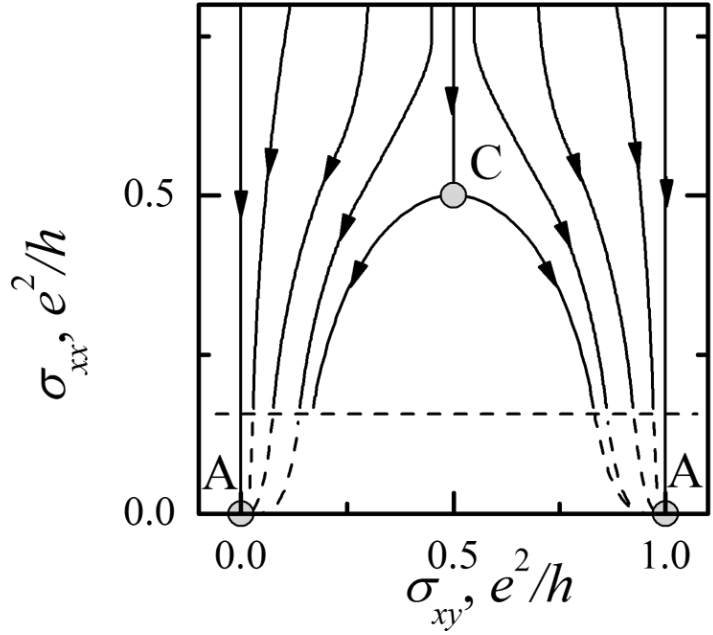

Fig. 1. Integral curves of the system of equations of two-parameter scaling according to the theory of Ref. 18. The arrows indicate the direction of motion of the $(\sigma_{xy}, \sigma_{xx})$ points as $L$ increases. The symbol **A** marks stable fixed points corresponding to plateaus of the IQHE; **C** is an unstable fixed point, corresponding to a delocalized state at the center of the Landau subband.

Explicit view of scaling diagrams in coordinates $(\sigma_{xy}, \sigma_{xx})$ was obtained by Khmel'nitskii [19] from the general properties of the symmetry and periodicity of the functions $\beta_{xx}$ and $\beta_{xy}$ in Eqs. (1.4). Fig. 2. illustrates the phase portrait of the system of equations (4).

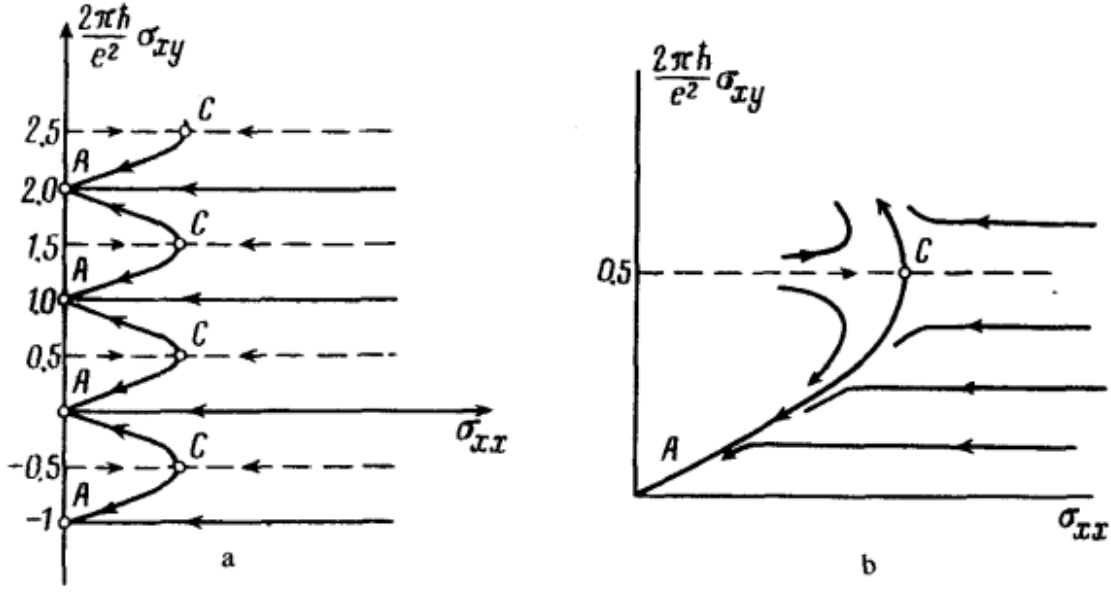

Fig. 2. Separatrices and fixed points (a) and integral curves (b) of the system of renormalization-group equations (4) (after [19])

In [24], Pruisken et al. are considering the global scaling diagram of the conductance as well as the appearance of a critical fixed point in strong coupling regime. For this fixed point theory predicts a metallic phase at the Landau band center as well as the following scaling result for the conductivities:

$$\sigma_{ij}(L,B) = g_{ij}([L/\xi]^{1/\gamma}), \qquad (9)$$
$$\xi \sim |B - B_c|^{-\gamma},$$

here, the function $g_{ij}(X)$ is a regular function of its argument, $B_c$ is the critical magnetic field strength and $\gamma$ stand for the critical index of the localization length $\xi$. The extensive numerical work on the free electron gas has been performed and the quoted best value for the critical index is $\gamma = 2.3$ for a short-range disorder potential [25].

In particular, the authors of [24] introduce a two- dimensional network of edge states and tunneling centers (saddle points) as a model for smooth disorder. This network is then used to derive a mean field theory of the conductance and they work out the characteristic temperature scale at which the transport crosses over from mean field behavior at high $T$ to the critical behavior plateau transitions at much lower $T$ (see inset in Fig. 3).

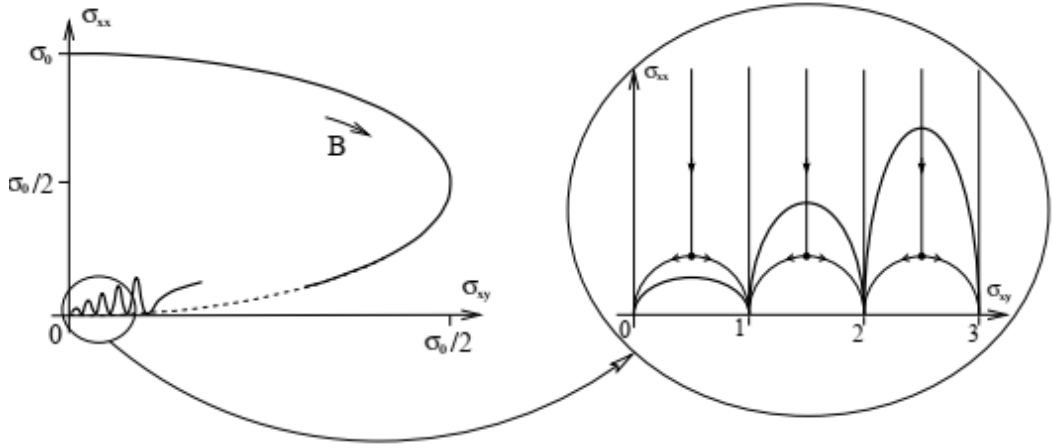

Fig. 3. Sketch of the mean field conductance for a short-range disorder potential. The inset is the strong field limit or quantum Hall regime. The renormalization group flow lines indicate how the mean field theory results change after successive length scale transformations (after [24]).

For experimental verification of the (two-parameter) renormalization-group (RG) theory of the integer quantum Hall effect, the method of temperature-dependent scaling has been used since the initial work of Wei et al. (1985) [20]. The conceptual translation of length scale $L$ into $T$ is based on the thought of Thouless length $L_\varphi$ as the effective sample size at finite temperatures. Since the phase coherence length $L_\varphi \sim T^{-p/2}$ ($p$ being the temperature exponent of the inelastic scattering rate), the limit of $L \to \infty$ ($L_\varphi \to \infty$) corresponds to the limit of $T \to 0$.

In order to carry out direct comparison with the RG flow diagram, the measured $\rho_{xx}(B,T)$ and $\rho_{xy}(B,T)$ are converted into $\sigma_{xx}(B,T)$ and $\sigma_{xy}(B,T)$. The results are plotted as a series of $\sigma_{xx}(\sigma_{xy})$ dependences at various temperatures with the $T$-driven flow lines at fixed $B$ and the limit $T \to 0$ is analyzed.

To analyze the temperature dependence of the conductivity $\sigma_{xx}$, it is convenient to start from the equation [20]:

$$\sigma_{xx}(T) = -\int dE \frac{\partial f(E-E_F)}{\partial E} \sigma(E), \tag{10}$$

where $f(E - E_F)$ is the Fermi-Dirac distribution function, and $\sigma(E)$ is the partial contribution to the dissipative conductivity of the states with energy $E$. Since only delocalized states in the energy interval $|E - E_c| \leq \Gamma$ (where $\Gamma$ is the non-temperature broadening of Landau levels, $E_c$ is the edge of mobility) contribute to the conductivity in the QHE regime, we can write the partial conductivity as:

$$\sigma(E) = \sigma_c \frac{\Gamma^2}{(E-E_c)^2+\Gamma^2}. \tag{11}$$

When $E_F = E_c$ we find from (10) and (11) that

$$\sigma_{xx}(T) = \frac{\pi}{4}\sigma_c \frac{\Gamma}{kT} \; (kT > \Gamma) \\ \sigma_{xx}(T) = \sigma_c (kT < \Gamma). \tag{12}$$

The quantity $\sigma_c = \sigma(E_c)$ in Eqs. (11) and (12) at zero temperature depends only on the linear size $L$ of 2D system. This dependence is determined by the system of equations of two-parameter scaling, Eqs. (4), which for $\sigma_{xy} = i + 1/2$ reduces to two independent equations where the symbol $\sigma^*$ corresponds to the zero of the function $\beta_{xx}$: $\beta_{xx}(i + {}^1/_2, \sigma^*) = 0$ (see Eq. (8))

Two regions can therefore be distinguished in the temperature dependence of the peak amplitude of $\sigma_{xx}(T)$ ($\sigma_{xx}^{max}$). In low-temperature regions, $kT < \Gamma$, the scaling regime realized, in which the temperature dependence $\sigma_{xx}^{max}(T)$ is completely determined by the Thouless length $L_\varphi$ and, as the analysis shows, $\sigma_{xx}^{max}$ *decreases with lowering of T*.

The limiting value of $\sigma^*$ corresponds to $T = 0$ in an infinite sample. It is just the decrease of $\sigma_{xx}^{max}(T)$ with decreasing temperature that serves as an empirical criterion for the transition to the scaling regime in a real sample.

When $kT > \Gamma$, the scaling dependence does not hold, the thermal smearing of the Fermi step $kT$ becomes the main factor and $\sigma_{xx}^{max}$ *increases with T lowering*. The results explain the apparent lack of scaling which may be seen in the transport data taken from arbitrary samples at finite $T$. The next section is devoted to the experimental study of temperature-dependent scaling diagrams for the particular 2D structures.

**4. Scaling diagrams (experiment)**

The conductivity components $\sigma_{xx}$ and $\sigma_{xy}$ is usually obtained by inverting the resistivity tensor:

$$\sigma_{xx} = \frac{\rho_{xx}}{\rho_{xx}^2 + \rho_{xy}^2},$$
$$\sigma_{xy} = \frac{\rho_{xy}}{\rho_{xx}^2 + \rho_{xy}^2}, \tag{13}$$

where $\rho_{xx}(B,T)$ and $\rho_{xy}(B,T)$ are the "raw" experimental data. The scaling diagrams for conductance are then created as the graphs of the dependences of $\sigma_{xx}$ on $\sigma_{xy}$.

In this subsection we provide the examples of constructing the scaling diagrams for conductivities from the experiments on different 2D materials. The data from some early and recent works are presented selectively with an emphasis on our works.

The overlook of early (pre-1995) experimental works on the scaling diagrams can be found in Huckestein's review [16]. Here, a brief historical outline on experimental studies of RG diagrams from "naive" pictures for Si – MOSFET and Ge/GeSi heterostructures to high-quality diagrams for InGaAs – based structures and, further, to sophisticated graphs for HgTe DQW and monolayer graphene is presented.

In [20], the experimental study of the quantum Hall effect in **In$_x$Ga$_{1-x}$As/InP** heterostructures is presented. Detailed measurements were made of the temperature and magnetic field dependences in the electronic transport coefficients, $\sigma_{xx}$ and $\sigma_{xy}$, for the *n* = 0 and 1 Landau levels. The results were studied in the context of the (two-parameter) renormalization-group theory of the integer quantum Hall effect (see Fig. 4).

In [21], the scaling relation of the diagonal and the Hall conductivities is experimentally studied in the (0 ↓ +) Landau level (0 refers to Landau quantum number, ↓ to antiparallel spin to the magnetic field, (+, - ) correspond to lower and higher valley) of an **n-channel inversion layer on a silicon (001) surface** in magnetic fields of 13 T and 15 T at temperatures between 1.5 K and 4.2 K.

Kawaji et al. [21] plot $\sigma_{xx}$ against $\sigma_{xy}$ in the lower half of the (0 ↓ +) level in two magnetic fields, 15 T and 13 T, for several fixed gate voltages at eight different temperatures between 1.5 K and 4.2 K. The result is shown in Fig. 5 in two different magnetic fields. At the low temperature limit, the results show a single flow line in the $\sigma_{xx}$ and $\sigma_{xy}$ space which support Ando's result [26].

In [27], it was shown experimentally that the magnetoconductivity of **Si – MOSFETs** (silicon metal-oxide-semiconductor field-effect transistors) at *T*<1 K is not described by the kinetic equation, which predicts that the maxima in $\sigma_{xx}$ will increase linearly with the index of the Landau level. The results can be explained on the basis of a two-parameter scaling theory.

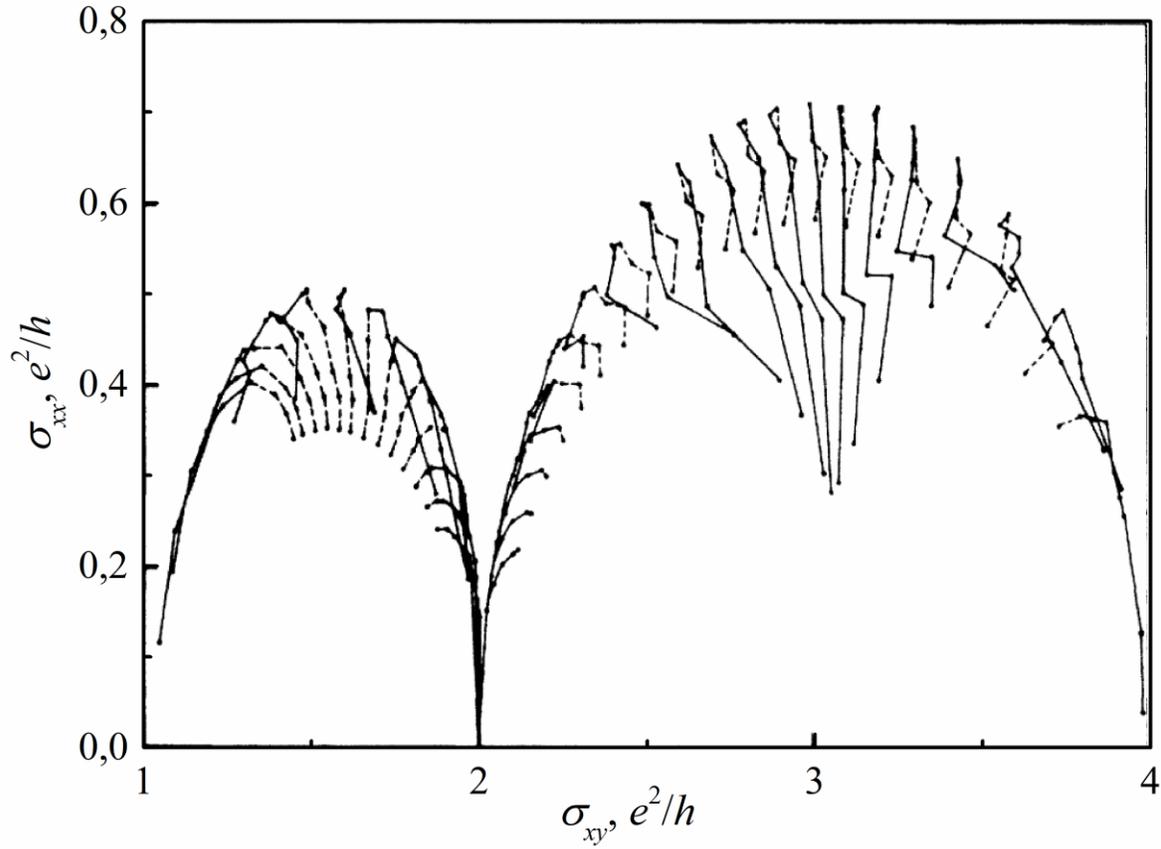

Fig. 4. Experimental $\sigma_{xx}(T)$ and $\sigma_{xy}(T)$ plotted as the $T$-driven flow lines from $T = 10$ to 0.5 K. The dashed lines are from 10 to 4.2 K and the solid lines from 4.2 to 5 K. Note the "flow" of points $(\sigma_{xy}, \sigma_{xx})$ to fixed point (3, 0) at $T < 4.2$K that corresponds to the formation of the $i = 3$ Hall plateau with decreasing temperature (after [20]).

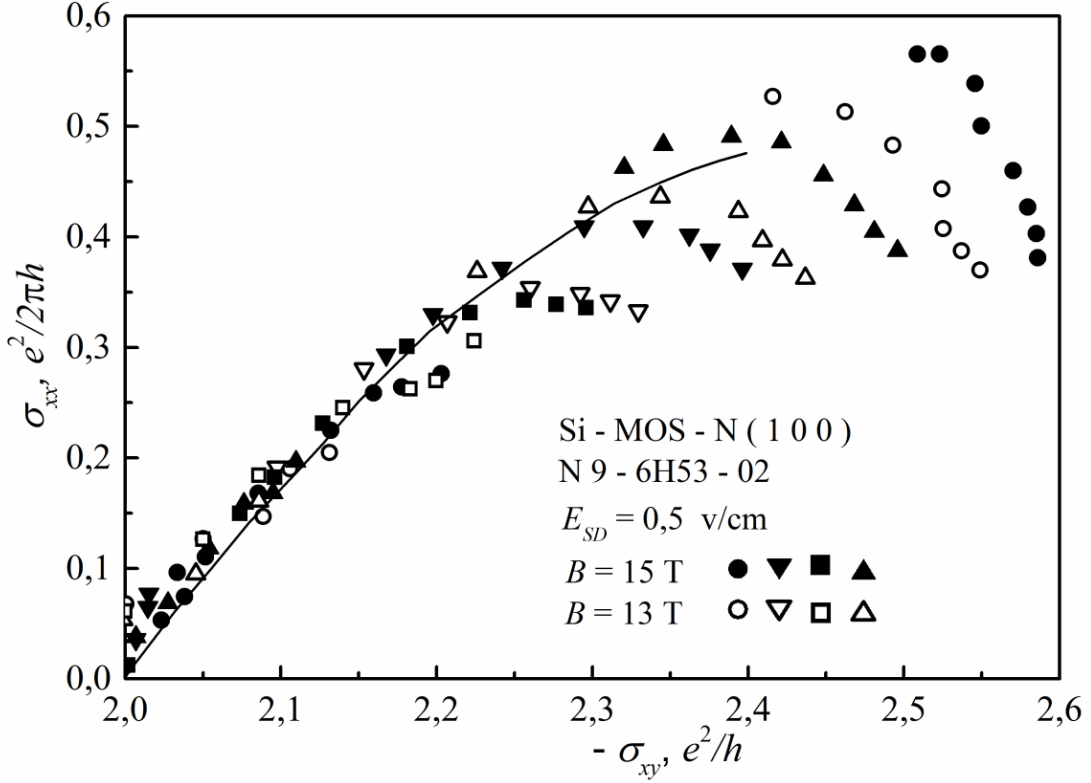

Fig. 5. Flow of diagonal conductivity $\sigma_{xx}$ vs Hall conductivity $\sigma_{xy}$ in the lower half of the $(0\downarrow+)$ Landau level in an n-channel Si (001) MOSFET at temperatures from 4.2 K to 1.5 K in magnetic fields of 13 T and 15 T. The solid line is calculated by use of an approximate expression to Ando's numerical scaling relation [26] (after [21]).

Fig. 6. shows phase paths for the $0\uparrow-$ and $1\downarrow+$ sublevels; these paths are typical of all the sublevels respectively the zeroth (N = 0) and first (N = 1) Landau levels. It follows from the theory of Refs. 18 and 19 that $\sigma_{xy}$ varies along the paths (the arrows show the direction of increasing $L_\varphi$) toward points of integer quantization. A decrease in $\sigma_{xx}$ at $T < 0.6$ K for N = 0 and at $T < 2$ K for N = 1 occurs for all initial values of $\sigma_{xy}$, including values which are approximately half-integers. The result of the present paper supports the conclusion of Refs. 18 and 19 that there is a set of phase paths in the $(\sigma_{xx}, \sigma_{xy})$ plane. At large values of $\sigma_{xx}$, the paths run nearly parallel to the ordinate axis, in agreement with Ref. 19.

Thus, in [27], it has been shown that at sufficiently low temperatures in a strong magnetic field the magnetoconductivity of silicon MOSFET structures is not described by a kinetic equation. As the temperature is lowered, quantum-mechanical effects come into play, causing the maxima of the diagonal conductivity for the zeroth and first Landau levels to move closer together. The results can be explained on the basis of a two-parameter scaling theory.

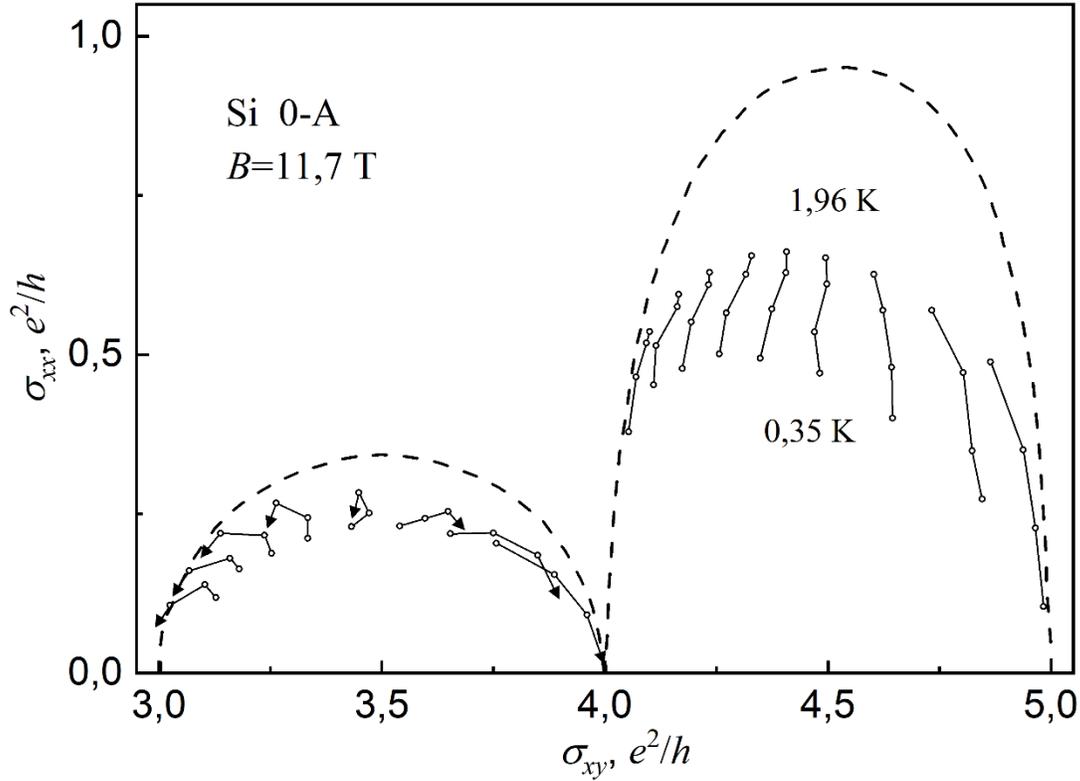

Fig. 6. Phase paths in the $\sigma_{xx} - \sigma_{xy}$ plane for sample Si-0-A at $H = 11.7$ T in the temperature interval 1.96–0.35 K (after [27]).

For multilayer **Ge/Ge$_{1-x}$ Si$_x$** ($x \cong 0.03$) heterostructures with two-dimensional $p$-type conductivity over the Ge layers, the temperature and magnetic dependences of the longitudinal resistivity $\rho_{xx}$ and the Hall resistivity $\rho_{xy}$ have been studied by Arapov et al. [28] in fields up to 12 T in the temperature interval of $T = (0.1\text{-}15)$ K.

The observed decrease of the amplitude of the $\rho_{xx}$ peaks with decreasing temperature for $T \leq 2$ K corresponds to a transition to the scaling regime under the conditions of the quantum Hall effect. Scaling diagrams in the ($\sigma_{xy}$, $\sigma_{xx}$) coordinates were constructed for the region of fields and temperatures of interest (see Fig. 7). It was found that, on the whole, the form of the diagrams corresponds to the theoretical predictions.

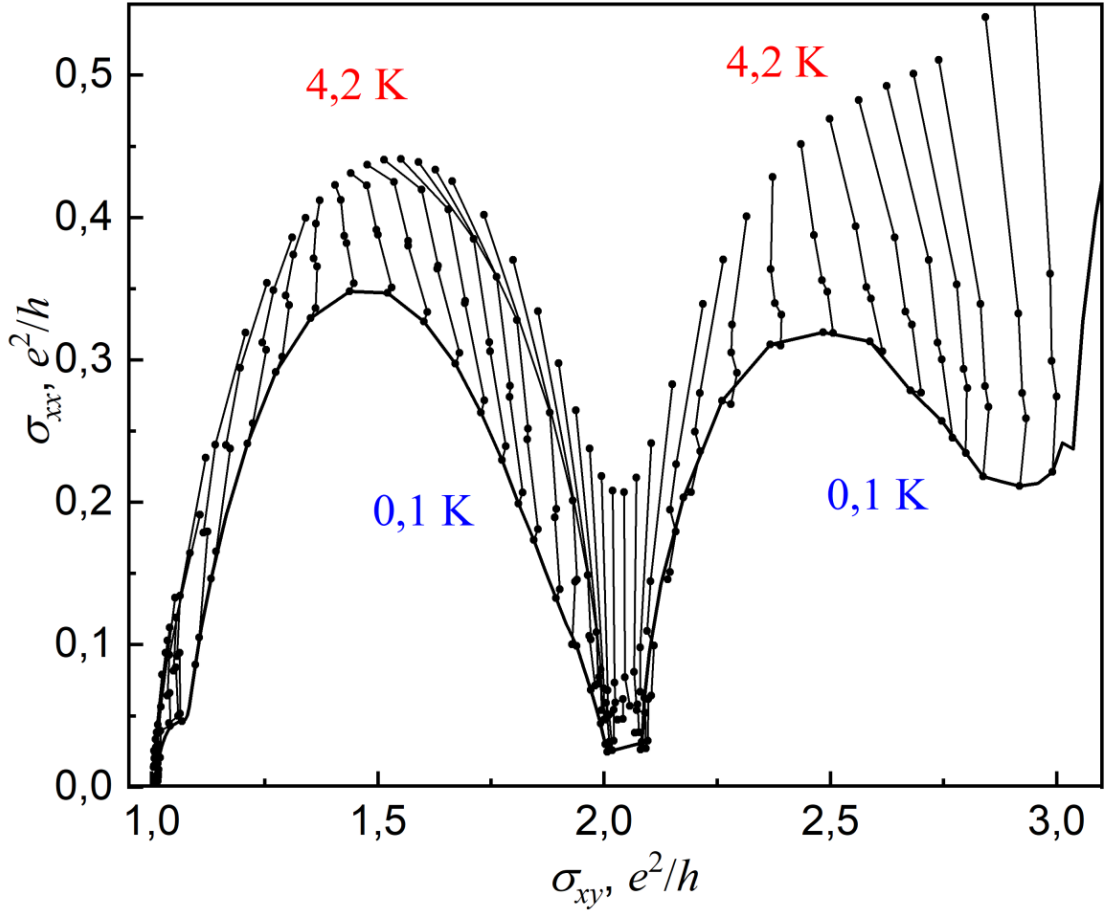

Fig. 7. Scaling diagram in ($\sigma_{xy}$, $\sigma_{xx}$) coordinates for Ge/Ge$_{1-x}$Si$_x$ sample. The flux lines are shown for fixed $B$ with a constant step of $\Delta B = 0.2$ T. The data are for $T =$ 0.1, 0.36, 0.9, 1.1, 1.7 and 4.2 K (after [28]).

The study of scaling diagrams in ($\sigma_{xy}$, $\sigma_{xx}$) coordinates for multilayer p-Ge/Ge$_{1-x}$Si$_x$ heterostructures confirms the main conclusion of the theory of two-parameter scaling: the presence of self-consistent variation of $\sigma_{xx}$ and $\sigma_{xy}$, as $L \to \infty (T \to 0)$. A consideration of the motion of the points on the ($\sigma_{xy}$, $\sigma_{xx}$) plane makes it possible to distinctly observe fixed points corresponding to the plateaus of the QHE (for i = 1, 2 and 3). An image of the plot of the dependence of $\sigma_{xy}$ on $\sigma_{xx}$ is an extremely sensitive method for studying the formation of the QHE plateaus as the external conditions change.

The longitudinal $\rho_{xx}(B)$ and Hall $\rho_{xy}(B)$ magnetoresistances are investigated in the integer quantum Hall effect regime in **GaAs/In$_{0.2}$Ga$_{0.8}$As/GaAs** double quantum well (DQW) nanostructures in the magnetic fields $B$ up to 16 T at temperatures $T = (0.05–4.2)$ K before and after IR (Infrared) illumination [29].

An evolution of the QHE picture can be traced by analyzing of fragments of $\sigma_{xx}$ vs. $\sigma_{xy}$ graphs for 1 ↑, 1 ↓ and 2 ↑, 2 ↓ peaks at $T$ from 4.2 K down to 0.05 K (Fig. 8 a,b). An initial rise in peak values with decreasing $T$ (4.2→ 1.6 → 1.0 K) is the usual consequence of the temperature dependence of the Fermi-Dirac distribution function when $kT$ is comparable to the Landau-level broadening and only a behavior in the range 1.0 - 0.05 K can be attributed to genuine scaling of the conductances [20].

It is seen from Fig. 8 that the spin splitting of the $N = 1$ Landau level at 4.2 K, 1.6 K and even at 1.0 K is not completely resolved. At lower $T$ there is a definite flow toward the $\sigma_{xy} = 3\,e^2/h$ fixed point, i.e., the formation of the i = 3 Hall plateau with complete spin splitting of 1 ↑ and 1 ↑ peaks. This corresponds physically both to exchange enhancement of the electron $g$-factor and to the narrowing of the bandwidth of delocalized states for 1 ↑ and 1 ↑ sublevels in the regime of scaling with decreasing $T$ [20].

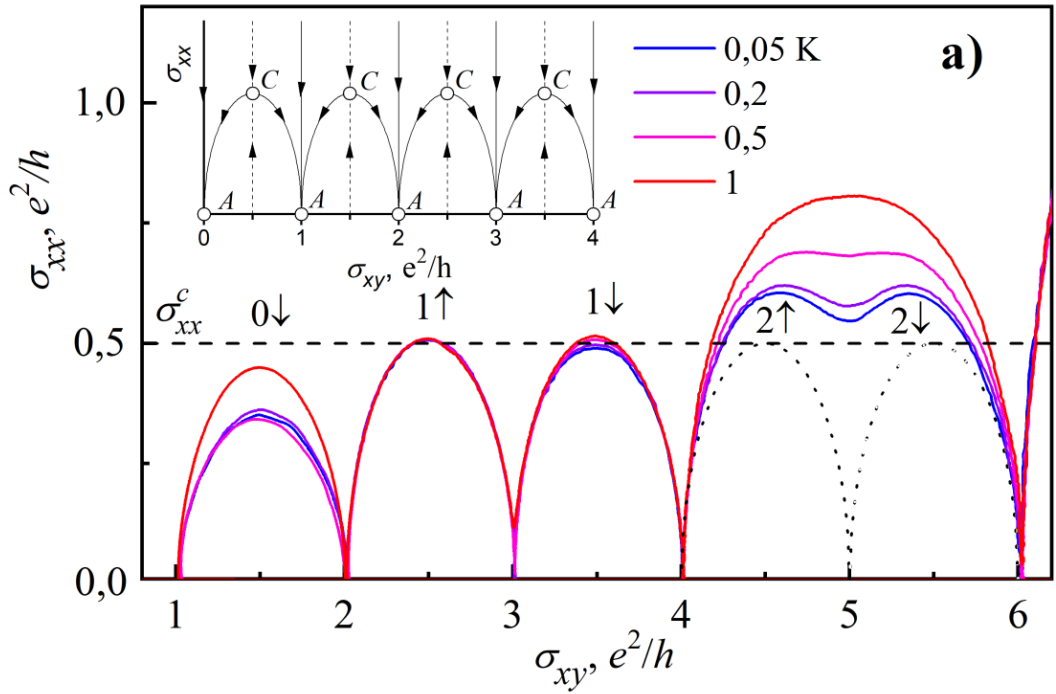

Fig. 8 (a) The graphs of the dependences σxx(σxy) for illuminated DQW sample at different temperatures (indicated in graph). Inset: Integral curves of the system of equations of two-parameter scaling according to the theoretical concepts (after [29]).

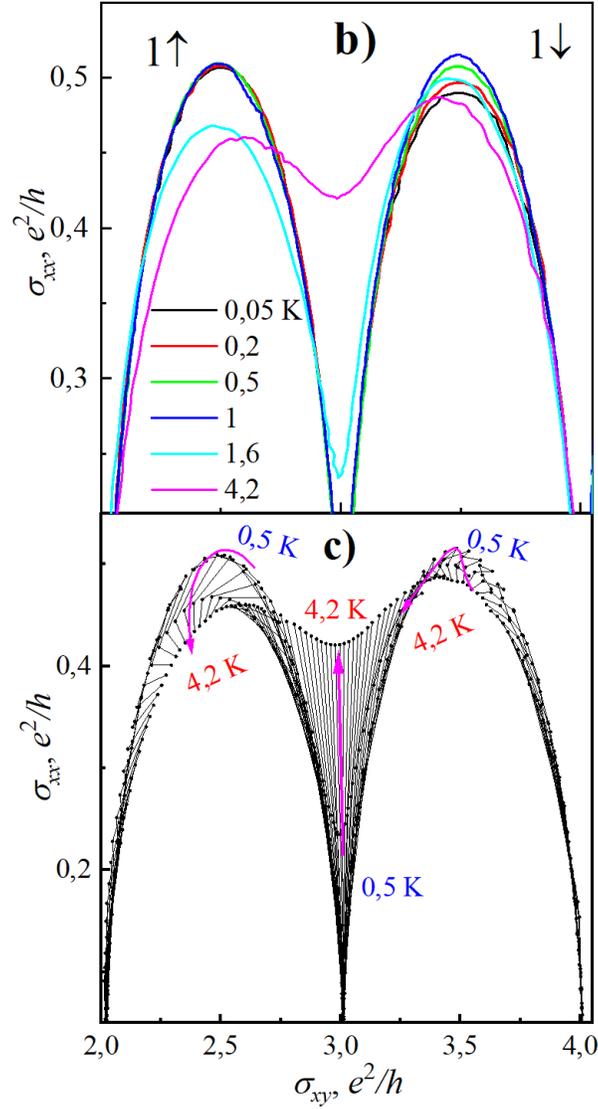

Fig. 8 (b) The detailed graphs of the dependences $\sigma_{xx}(\sigma_{xy})$ for illuminated DQW sample at different temperatures $T$ = 0.05-4.2 K (after [29]). (c) The same graphs as in 8(b) but the flow lines are given for fixed values of $B$, the data at $T$ = (0.5-4.2) K are used (new results). Arrows indicate an increase in temperature.

The spin splitting of the N = 2 Landau level is not completely resolved even at 0.05 K and thus the $i$ = 5 Hall plateau is only in the process of formation (Fig. 8b). A variant of the scaling diagram with the flow lines at fixed values of $B$ for the studied structure is presented on Fig. 8c.

In [30], the longitudinal and Hall resistivities in the quantum Hall effect regime at magnetic fields $B$ up to 9 T and temperatures $T$ = (1.8÷30) K for n-**In$_{0.85}$Ga$_{0.15}$As/In$_{0.82}$Al$_{0.18}$As** metamorphic nanoheterostructure were measured. In this work we present scaling diagrams for this sample at a wide range of Landau level (LL) filling factors, $\nu = (2 - 6)$ (Fig. 9).

The formation of RG separatrices is clearly visible in the Figs 9a,b with decrease of $T$. At the lowest temperature, $T = 0.4$ K, the value of $\sigma_{xx}$ at the vertices of the separatrices, $\sigma_{xx}^{max} = (0.33 \pm 0.01)\, e^2/h$, is practically independent of the LL number in accordance with the concepts of the two-parameter scaling hypothesis (see Fig. 2a and inset on Fig. 3).

Fig. 9b demonstrates the formation of a plateaus with $i = 2,3,4$ and a significant expansion of the region of magnetic fields, $\Delta B$, occupied by the plateau with $i = 3$ with decreasing $T$. It should be noted that for the samples we studied (InGaAs/GaAs [29] and InGaAs/ InAlAs [30]), the obtained RG diagrams are of fairly high quality.

In the communication [31], Pruisken et al. proposed universal scaling functions for the conductance parameters that are based on experimental data for the **plateau–insulator (PI) transition** in the quantum Hall regime taken from a low-mobility **InGaAs/ InP** heterostructure. The authors argue that PI transition in contrast to the plateau–plateau (PP) transition exhibits special features in order to obtain perfect scaling diagrams.

Thus, using the principle of particle–hole symmetry for the PI transition enables them to disentwine the intrinsic transport properties from the sample dependent effects of macroscopic inhomogeneities. Further, taking into account finite size scaling corrections leads to the following expressions for the "modernized" conductivities $\sigma_{xx} \to \sigma_0$ and $\sigma_{xy} \to \sigma_H$:

$$\sigma_0 = \frac{\rho_0}{\rho_0^2 + 1 + 2\eta\rho_0},$$
$$\sigma_H = \frac{1 + \eta\rho_0}{\rho_0^2 + 1 + 2\eta\rho_0}. \tag{14}$$

Here $\rho_0$ is a modernized (by taking into account the particle–hole symmetry) value of $\rho_{xx}$, and finite size scaling correction $\eta(T) = (T/T_1)^{y_\sigma}$ with $T_1 = (9.2\pm0.3)$ K and $y_\sigma = (2.43 \pm 0.08)$ being an irrelevant critical index.

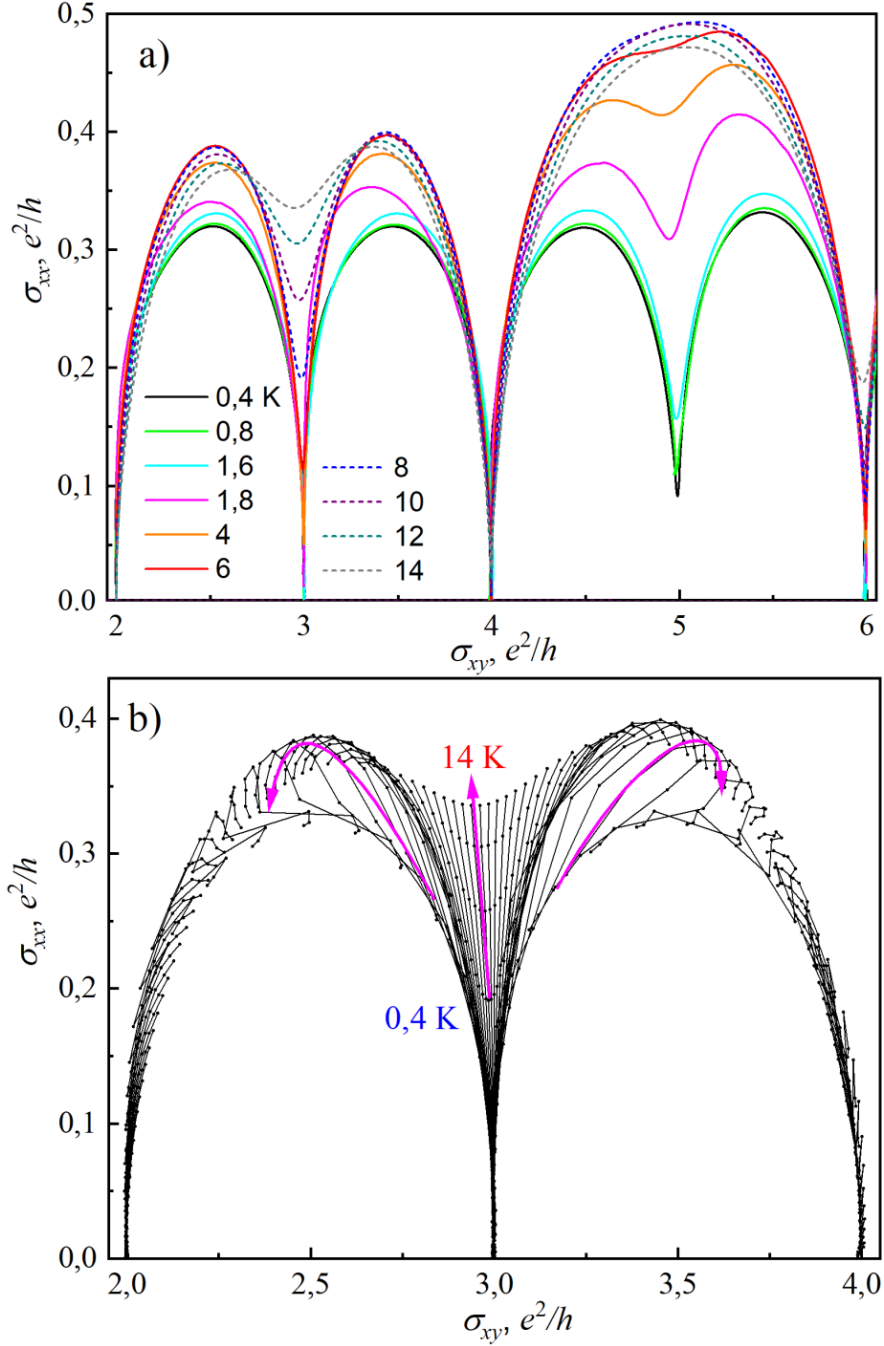

Fig. 9 (a)The graph of the dependences σxx(σxy) for SQW InGaAs/InAlAs sample with high InAs content for the transitions between QHE plateaus with numbers $i = 2, 3, 4, 5$ and $6$ at different temperatures. The lines are given for fixed temperatures $T = (0.4\text{-}14.0)$ K (indicated in graph). (b) The graph σxx(σxy) for transitions between QHE plateaus with $i = 2, 3, 4$, flow lines are given for fixed values of $B$, the data at $T = (0.4\text{-}14.0)$ K are used (new results).

When plotted the $T$-driven flow lines in the $\sigma_0, \sigma_H$ conductivity plane (Fig. 10) Pruisken et al. observed distinctly the features of universal scaling that previously remained concealed in the experiments on the PP transitions. In particular, the central flow line for $\nu = \nu_c = 0.5$ is clearly visible. Note that, in general, the scaling diagram in Fig. 10 demonstrates the flow of points $(\sigma_0, \sigma_H)$ as presented above in the left part of Fig. 2 from [19].

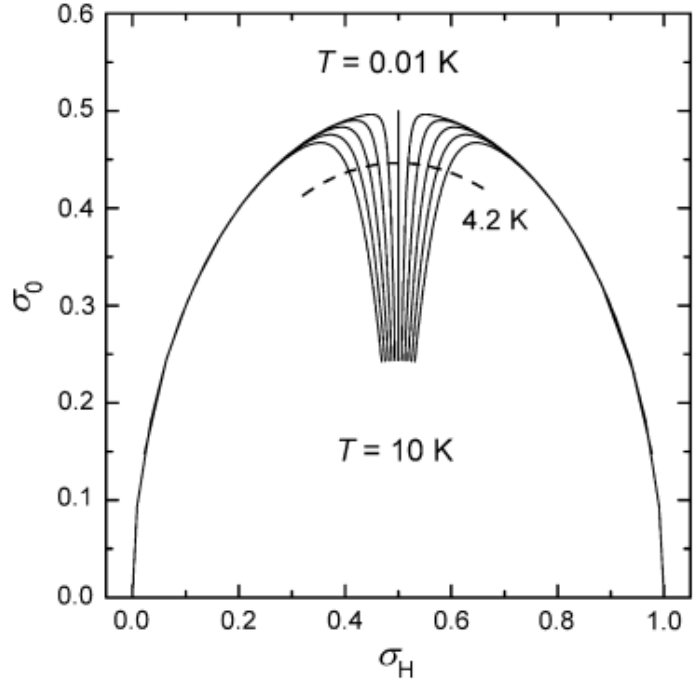

Fig. 10. Experimental $T$-driven flow lines with 0.01<$T$<10 K for different values of the filling fraction of the Landau band ($\nu$) near the critical value $\nu_c = 0.5$, according to Eq. (14) (after [31]).

Yakunin et al. [32, 33] report on the observation of an unconventional reentrant structure of the quantum Hall effect in a **p-type HgTe/Cd$_x$Hg$_{1-x}$Te double quantum well (DQW)** consisting of two HgTe layers of critical thickness, $d = d_c$, i.e., when a Dirac energy spectrum is formed in a single HgTe layer. The DQW consists of two HgTe layers with a thickness $d = 6.5 \pm 0.2$ nm separated by a $3 \pm 0.1$ nm Cd$_x$Hg$_{x-1}$Te barrier with $x = 0.71$.

The band dispersion of each of the two HgTe QWs taken separately is a Dirac cone, because the nominal thickness of the QWs corresponds to the critical thickness $d_c$ at which the gap disappears.

Calculations have shown that the DQW spectrum reproduce the so-called "bilayer graphene" (BG) phase.

Figs 11a,b show the magnetic field dependences of the Hall magnetoresistance for the sample under study at temperatures of (a) (1.8-12) K and (b) 0.26 K. One can see that in the region of magnetic fields of (1.5-2.0)T on the $\rho_{xy}(B)$ dependence, in place of the expected plateau–plateau 2–1 transition, a peak of the Hall MR is formed, which also manifests itself in the longitudinal MR in the form of a similar double peak $\rho_{xx}(B)$ in the same field.

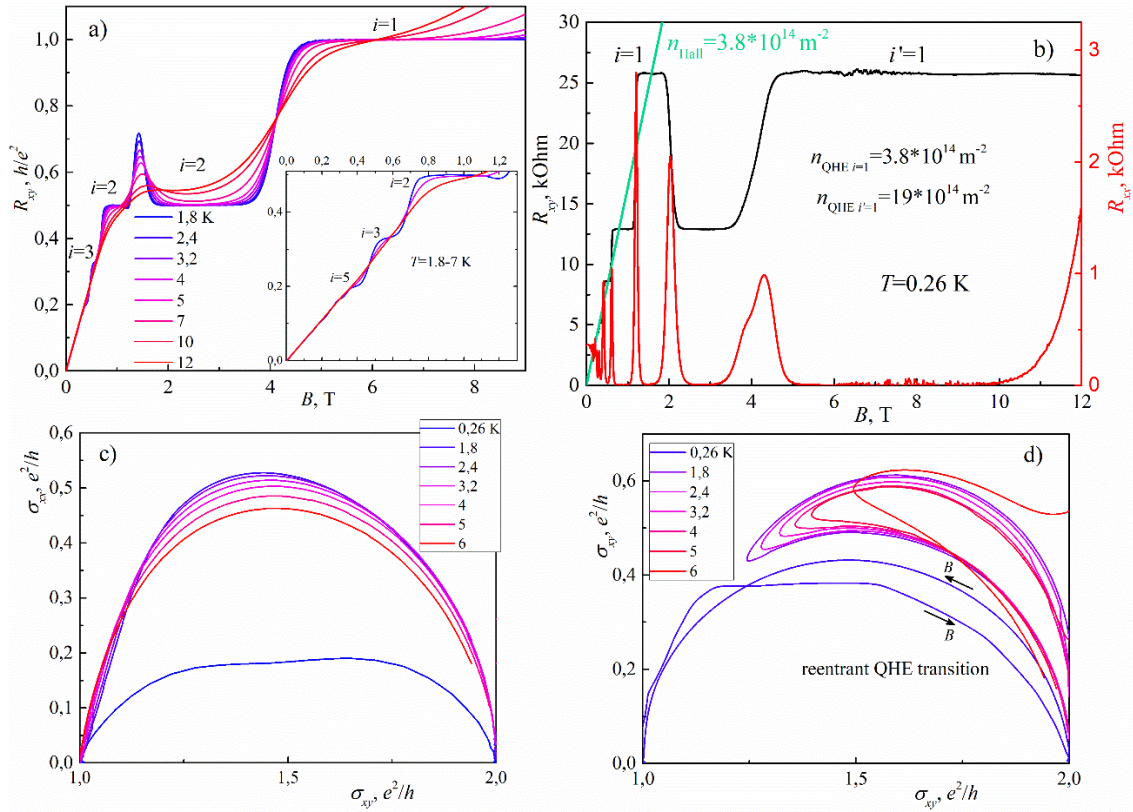

Fig. 11 (a) Dependences of the Hall, $R_{xy}$, component of the magnetoresistance tensor on the magnetic field $B$ in temperature range (1.8 – 12) K (after [32], [33]). (b) $\rho_{xx}(B)$ and $\rho_{xy}(B)$ dependences at $T = 0.26$ K (after [32, 33]). (c, d) Scaling diagrams for 2→1 transitions for the HgTe DQW sample with critical layer thickness. The lines are given for fixed temperatures $T = (0.26-6)$ K. The 2→1 transition at $B > 3$ T (c) and the reentrant 2→1→2 PPTs (d). The arrows indicate the direction in which the magnetic field $B$ increases (new results).

Thus, the observed QHE is a reentrant function of magnetic field between two $i = 2$ states (plateaus at $\rho_{xy} = h/ie^2$) separated by an intermediate $i = 1$ state in the shape of a peak placed on the long $i = 2$ plateau. When the temperature is reduced to $T = 0.26$ K, the formation of the plateau $i = 1$ occurs clearly, now the width of the plateau $i = 1$ is 0.7 T, however, in the field $B = 1.95$ T $\rho_{xy}(B)$ again returns to the values corresponding to the plateau $i = 2$ (Fig. 11b).

The anomalous $i = 1$ flat-top peak separates two different regimes: (i) a traditional QHE at relatively low fields corresponding to a small density of mobile holes, $p = 0.38 \cdot 10^{15} m^{-2}$, and (ii) a high-field QHE with a 2-1 plateau-plateau transition corresponding to a much larger $p = 1.9 \cdot 10^{15} m^{-2}$ (see Fig. 11b).

As claimed in [32, 33] the reentrant QHE revealed is an indication of switching between two valence states with different densities of holes in a structure with a BG-type spectrum: the state of high-mobility holes in the center of the Brillouin zone and the state of heavy holes in the lateral maximum of the valence subband.

Using the data for the longitudinal, $\rho_{xx}(B,T)$, and Hall, $\rho_{xy}(B,T)$, resistivities from the works [32, 33], we constructed the scaling diagrams $\sigma_{xx}$ vs $\sigma_{xy}$ at fixed temperatures both for the reentrant (2→1→2) transitions and for the high - field (2→1) PPT (Fig. 11 (c, d)).

At high-field (2→1) PPT, we see the traditional picture of $\sigma_{xx}$ vs $\sigma_{xy}$ dependences when the $\sigma_{xx}^{max}$ tends downwards with decreasing temperature for $T< 1.8$K. It correlates with the manifestation of the scaling regime for the temperature dependence of that transition width, $\Delta B \sim T^{\kappa}$, with the value of the critical index $\kappa = 0.37 \pm 0.03$, which is close to the universal value $\kappa = 0.42$ (see [34]).

The original view of the RG diagram for the double PPT at $B = (1.5 - 2)$ T reflects the unusual (reentrant) behavior of the QHE due to complex type of energy spectrum in HgTe DQW sample with critical layer thickness.

**Graphene**, which is a single layer of carbon atoms bonded in a honeycomb lattice, has continued to attract worldwide interest [35], [36].

In [37], Huang et al. reported the renormalization group (RG) flow for the transition between an insulator and the $\nu = 2$ QH state for strongly **disordered monolayer graphene** epitaxially grown on SiC, which, at zero magnetic field, becomes insulating as $T$ decreases. By changing the measurement temperature one can vary the effective sample size so as to study $T$-driven RG flow of a graphene device in the complex conductivity plane (scaling diagram for conductance).

Fig.12. shows the expected form of the RG diagram based on the semicircle law for the separatrix [18], [19], [38], [39]. For a system with a single conduction channel the semicircle separatrix represents a critical boundary for the $\nu = 2$ QH state, with a stable point on the $x$-axis at $(\sigma_{xy}, \sigma_{xx}) = (2e^2/h, 0)$. An unstable point at $(\sigma_{xy} = e^2/h, \sigma_{xx} = e^2/h)$ represents the boundary between two types of RG flow. To the left of the unstable point, the RG flows are towards $(\sigma_{xy} = 0, \sigma_{xx} = 0)$ which is the insulating state. To the right of the unstable point, the RG flows are towards $(\sigma_{xy} = 2\,e^2/h, \sigma_{xx} = 0)$, which is the $\nu = 2$ QH state.

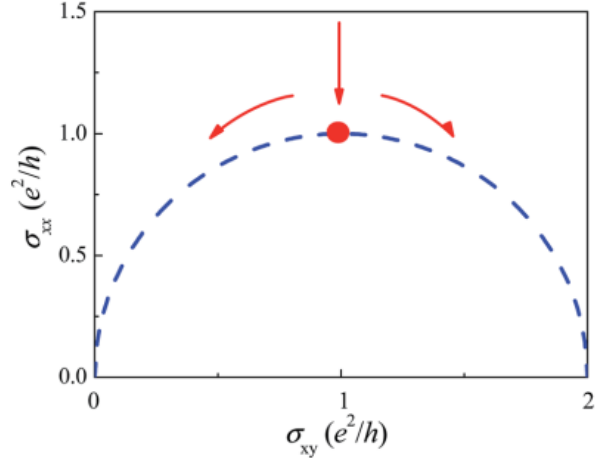

Fig.12. Schematic diagram showing RG flow for the 0–2 transition in the conductivity plane. The red dot corresponds to the unstable point. The red arrows indicate the directions of the RG flow (after [37]).

However, the scaling (RG flow) diagram of $(\sigma_{xy}, \sigma_{xx})$ that was actually constructed from experimental data turned out to be quite unusual (Fig. 13 (a) and (b)). At low $B$ the $T$-driven flow for fixed $B$ forms a set of curves that, at the lowest $T$, converge toward the insulating state at (0,0). Such results strongly demonstrate that the device behaves as an insulator.

At $B > 0.3$ T, the curves for fixed $B$ bend outward in the range near $\sigma_{xx} = e^2/h$ and at higher $B$ a cusp appears near the unstable point at $(e^2/h, e^2/h)$. The authors emphasize that such cusp-like RG flow is a unique feature in strongly disordered graphene.

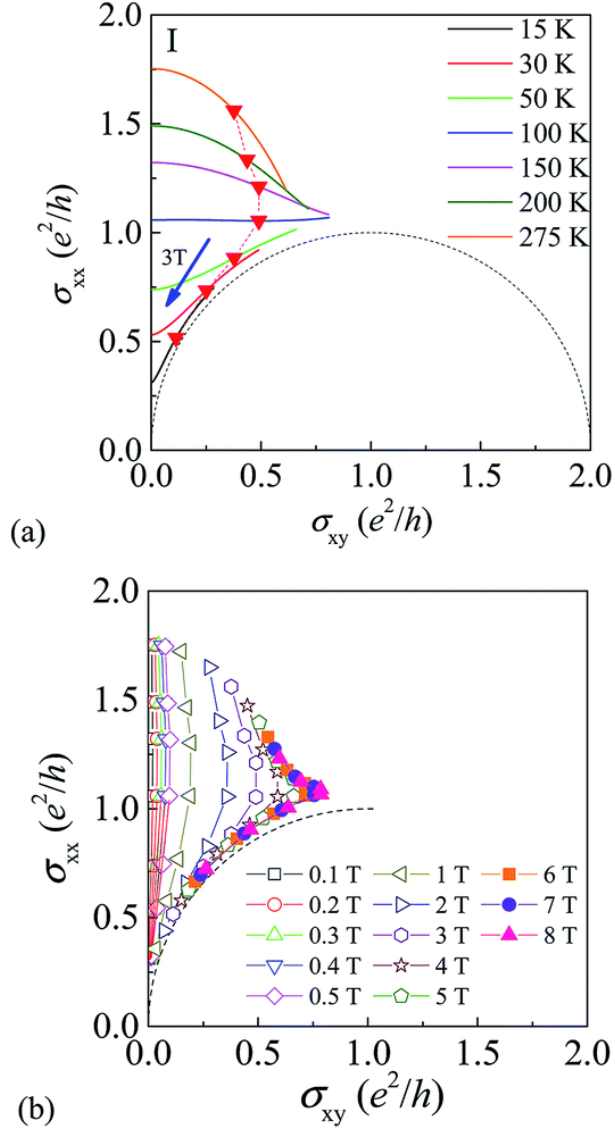

Fig. 13. (a) Conductivity $\sigma_{xx}$ plotted against $\sigma_{xy}$. The dotted curves denote the theoretical prediction of semicircle $\sigma_{xx}$ vs $\sigma_{xy}$ relation for the 0–2 transition. Each group of triangle markers connected by dashed lines denotes the data for the same magnetic field ($B = 3$ T). The arrows indicate the flow line to the low temperature extreme at fixed magnetic fields. (b) Detailed RG flow over a wide range of $T$ (after [37]).

A possible reason for the observed cusp-like RG flow is that the $\nu = 2$ QH state in graphene on SiC is very robust so that even strong disorder and high $\rho_{xx}$ within the device can co-exist with the presence of a QH-plateau-like structure in $\rho_{xy}$. These new results cannot be explained by any existing RG models and further experimental and theoretical studies are needed in order to probe the unexplored area in the field of disordered 2D materials.

Recent quantum Hall experiments in the **high-quality single graphene layers** [35], [36] have produced spectacular results that are markedly different from the conventional quantum Hall effect in semiconducting layers. In [39], the authors argue that these experimental results, *at least for intermediate values of the external magnetic field,* are compatible with a suggestion that a particular modular group (of $\Gamma_\theta$ symmetry) might be a QHE symmetry in monolayer graphene.

The theoretical analysis within the framework of this RG symmetry led to a series of predictions for high-mobility graphene samples: predictions for the locations of the QH plateaus, the positions of critical points on PP transitions and a semicircle law for conductivities in the vicinity of these transitions.

The consequences of $\Gamma_\theta$ symmetry for the flow of the conductivities are given on Fig. 14: it is a detailed version of the resulting RG flow diagram both for the integer and for the fractional quantum Hall plateaus in the high-mobility monolayer graphene.

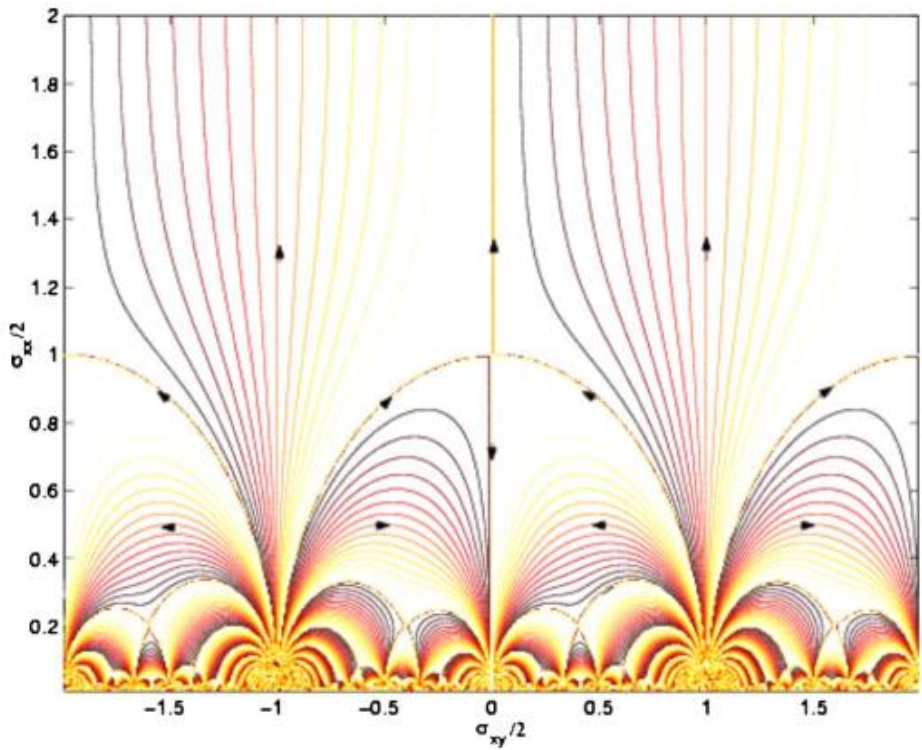

Fig. 14. RG flow diagram for $\Gamma_\theta$ symmetry in high-mobility monolayer graphene, the flow directions shown by arrows correspond to *increasing* temperature (after [39]).

In [39], the experimental data of Ref. 35 have been examined as a preliminary test (see Fig. 15). This figure plots the Ohmic resistivity $\rho_{xx}$ and the Hall conductivity $\sigma_{xy}$, as reported in Ref. 35, as a function of carrier density $n$.

By assuming semicircular trajectories for the PP transitions the authors obtain the blue curves in Fig. 15, overlaid on the experimental curves. The visual agreement between the semicircular curves and the experimental data seems rather good for intermediate QHE plateau numbers and worse for the higher plateaus. The authors propose to continue an extension of RG calculations, as well as a comparison of these results with new experimental data for IQHE and FQHE in graphene.

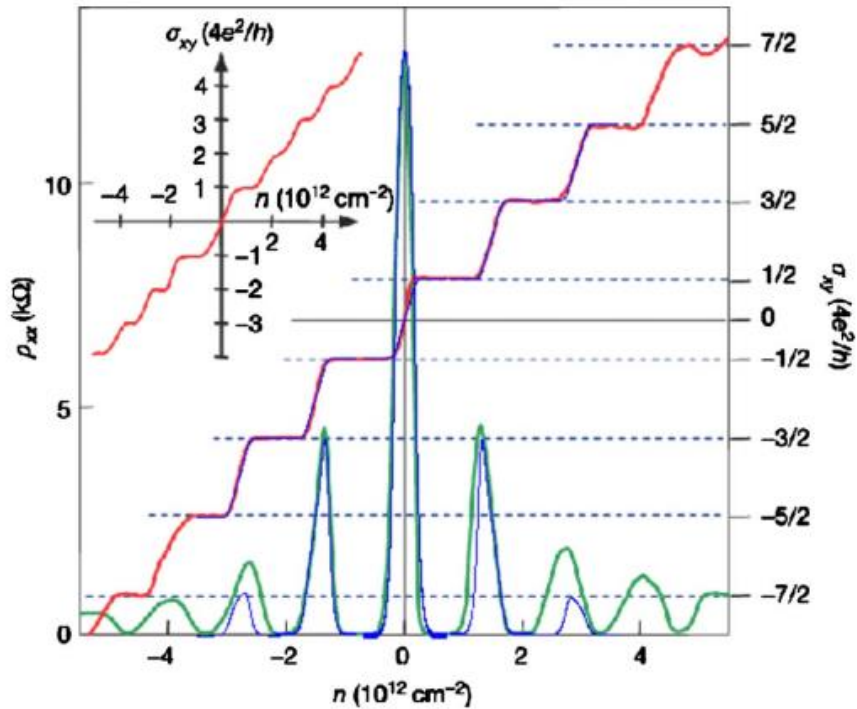

Fig. 15. A semicircle (solid blue line) overlaid on the data of Ref. 35 (after [39]).

## 5. Discussion and conclusions

A detailed analysis of the physical foundations for the two-parameter scaling hypothesis was carried out by Gantmakher in monograph [3] (Chapter 9, paragraph 9.7). On the other hand, Huckestein [16] emphasizes the experimental difficulty to obtain reliable flow-diagrams close to the transitions at half-integer Hall conductivity. As the temperature diminished the transition regions become very narrow and inhomogeneities in the electron density will lead to irregular behavior of σ$_{xx}$ on σ$_{xy}$ dependencies near the transition.

Let us compare the available experimental graphs $\sigma_{xx}$ vs $\sigma_{xy}$ for different materials with the theoretical predictions for the scaling diagrams of conductance [18], [19] (see Figs 1 and 2).

In general, the dependences σ$_{xx}$ vs σ$_{xy}$ obtained from the experiment by inverting the resistivity tensor $\rho_{ij}(B,T)$ are far from ideal, they are often very asymmetrical with respect to half-integer values of σ$_{xy}$. In this work, we have selected only sufficiently symmetrical examples. Thus, the scaling diagrams of good quality are presented by Wei et al. [20] for InGaAs/InP (Fig. 4), by Kravchenko and Pudalov [27] for Si – MOSFET (Fig. 6), as well as by Arapov et al. [29] and Gudina et al. [30] for InGaAs QWs (Figs. 8 and 9).

The following achievements should be noted:

- in the best examples, at low temperatures the formation of separatrices $\sigma_{xx}$ vs $\sigma_{xy}$ on the graphs of the RG diagrams is well expressed, at that the value of $\sigma_{xx}^{max}$ at the vertices of the separatrices is practically independent of the Landau level number, in accordance with the concepts of the two-parameter scaling hypothesis.

- it can be clearly observed the movement of flow lines to the stable fixed points **A** corresponding to the QHE plateaus, see Figs 4, 8 and 9 for InGaAs systems ([20], [29], [30]) and Fig. 7 for multilayer Ge/GeSi heterostructures (Arapov et al. [28]). An image of the plot of the dependence of $\sigma_{xx}$ on $\sigma_{xy}$ is an extremely sensitive method for studying the formation of the QHE plateaus with decreasing temperature;

- note also that a correlation takes place: when the universal critical index $\kappa \approx 0.42$ (or $\kappa \approx 0.21$) is observed in the *T* dependence of the PP transition width, the flow lines *always* (in all the examples above) move *downwards* with decreasing temperature on the σ$_{xx}$ vs σ$_{xy}$ diagram, i.e. (according to theory) the condition *kT*< Γ is satisfied.

The main difficulty is to see in experiment the motion of ($\sigma_{xy}$, $\sigma_{xx}$) points to an unstable fixed point **C**, corresponding to a delocalized state at the center of Landau subband. This central integer curve corresponds to a single (at *T* = 0) energy when the electron trajectory passes through the saddle points (tunneling processes) and finite sample size scaling corrections must be taken into account (for more information on finite-size scaling see, for example, the review by Huckestein [16]).

The symmetrical scaling diagram of the conductance parameters with a pronounced flow line at half-integer Hall conductivity (for the critical value of $\nu = \nu_c = 0.5$) Pruisken et al. [31] were able to get only after using the principle of particle–hole symmetry and taking into account finite size scaling corrections (see Eq. (14)) for the plateau–insulator transition in the QHE regime at InGaAs/InP heterostructure (see Fig. 10).

To avoid this problem Wei et al. [40] have studied the critical behavior of each component of the resistivity tensor separately: the temperature dependence of the maximum $(d\rho_{xy}/dB)$ and the inverse width $(\Delta B)^{-1}$ of the $\rho_{xx}$ peaks is plotted for several Landau levels in InGaAs/InP heterostructure. The slope of the straight lines gives $d\rho_{xy}/dB|_{max} \sim T^{-\kappa}$ and $\Delta B \sim T^{\kappa}$ with $\kappa = 0.42$.

In current literature, this version of the scaling hypothesis, which takes as a basis the critical behavior of a system (the divergence of localization length) in the centers of Landau levels, dominates (see review [16] and references therein). Thus, the integer quantum Hall effect regime can be considered as a sequence of the localization–delocalization–localization quantum phase transitions (a special case of an insulator–metal–insulator transition) when the density of states of a disordered 2D system is scanned by the Fermi level in a quantizing magnetic field.

At finite temperatures the width of the transition between neighboring IQHE plateaus, as well as the width of the corresponding peak in the magnetic-field dependence $\rho_{xx}(B)$ should tend to zero by the power-law dependence $T^{\kappa}$, where $\kappa = p/2\gamma$, $\gamma$ is the critical exponent of the localization length and the exponent $p$ depends on the inelastic scattering mechanism. The value of critical exponent κ = 0.42 is compatible with a numerical short-ranged potential value γ = 2.3 [16], [25] for the Fermi-liquid electron-electron interaction exponent $p = 2$.

Thus, one can say that the IQHE regime is a manifestation at finite $T$ of the universal zero–temperature quantum phase transitions. This result lends itself well to experimental verification, and the key question of the universality (or non-universality) of the observed $\kappa$ values is still being discussed (see the overview of the problem, for example, in [41] or in [42]).

Recently, the method of constructing scaling diagram for conductance parameters (RG flow diagram) has received new life in the study of quantum phase transitions in the QHE regime for monolayer graphene. Such method provides important information on the insulating behavior in monolayer epitaxial graphene grown on SiC [37] and on the transition between an insulator and the $\nu = 2$ QH state. RG flow lines in this disordered two-dimensional system are understood by referring to the semicircle law for separatrix with a clear unstable point at $\nu = \nu_c = 1$.

In [37], Huang et al. argue that the unique insulator-relativistic QH transition in disordered graphene is better studied by the RG flows rather than by the crossing point in $\varrho_{xx}$ since RG flows described by the semicircle law strongly point to the floating-up of the $N = 0$ Landau level at low magnetic fields. Moreover, the pronounced unstable fixed point in the RG flow diagram unequivocally shows signature of a quantum phase transition.

In [39], it has been argued that the recently observed quantum Hall effect [35, 36] in the high-quality monolayer graphene to be governed by emergent modular RG symmetry ($\Gamma_\theta$ symmetry) whose predictions differs from those obtained for the conventional quantum Hall effect in semiconductors. Within the framework of $\Gamma_\theta$ symmetry, a large series of flow RG diagrams with special semicircle laws for separatrices is calculated for both the integer and the fractional quantum Hall plateaus in the high-mobility monolayer graphene.

**Acknowledgments**